\documentclass[runningheads]{llncs}
\usepackage{graphicx}
\usepackage{cite}
\usepackage{amsmath,amssymb,amsfonts}
\usepackage{algorithmic}
\usepackage{graphicx}
\usepackage{textcomp}
\usepackage{xcolor}
\usepackage{multirow}
\usepackage{tabularx}
\usepackage{threeparttable}
\usepackage{placeins}
\usepackage{footmisc}
\usepackage{url}
\usepackage[numbers]{natbib}
\begin{document}
\title{Are Fact-Checking Tools Helpful? An Exploration of the Usability of Google Fact Check}
%
%

\author{Qiangeng Yang\inst{1}\and
Tess Christensen\inst{1} \and
Shlok Gilda\inst{1} \and
Juliana Fernandes\inst{1} \and
Daniela Oliveira\inst{2} \and
Ronald Wilson\inst{1} \and
Damon Woodard\inst{1}
}

\authorrunning{Q. Yang et al.}

\institute{University of Florida, Gainesville, FL 32611, USA\\
\email{q.yang@ufl.edu}\and
National Science Foundation, Alexandria, VA 22314, USA}

\maketitle              
\begin{abstract}
Fact-checking-specific search tools such as Google Fact Check are a promising way to combat misinformation on social media, especially during events bringing significant social influence, such as the COVID-19 pandemic and the U.S. presidential elections. However, the usability of such an approach has not been thoroughly studied. We evaluated the performance of Google Fact Check by analyzing the retrieved fact-checking results regarding 1,000 COVID-19-related false claims and found it able to retrieve the fact-checking results for 15.8\% of the input claims, and the rendered results are relatively reliable. We also found that the false claims receiving different fact-checking verdicts (i.e., ``False,'' ``Partly False,'' ``True,'' and ``Unratable'') tend to reflect diverse emotional tones, and fact-checking sources tend to check the claims in different lengths and using dictionary words to various extents. Claim variations addressing the same issue yet described differently are likely to retrieve distinct fact-checking results. We suggest that the quantities of the retrieved fact-checking results could be optimized and that slightly adjusting input wording may be the best practice for users to retrieve more useful information. This study aims to contribute to the understanding of state-of-the-art fact-checking tools and information integrity. 

\end{abstract}
\section{Introduction}
Fact-checking has been one of the most common ways to combat misinformation on social media in recent years \cite{vinhas2022fact}. It is a significant approach to protect information integrity under the growing influence of social media where people can freely share information and potentially disseminate misinformation \cite{allcott2019trends}. Several mainstream social media platforms such as Facebook \cite{facebook_about_2021} and YouTube \cite{google_topical_2021} display labels disclosing potential inaccuracy or context to foster transparency by collaborating with fact-checking organizations that regularly detect suspicious claims, compile evidence, and publish fact-checking reports. Traditional manual scrutiny usually suffers potential delays and the gap between public focus and fact-checking efforts while facing fast and large-scale dissemination of online information \cite{ribeiro2021can}, regarding which researchers proposed automated fact-checking frameworks to collect existing fact-checking results or related evidence for real-time review \cite{guo2022survey}. 

Neither manual nor automatic approaches can guarantee all false claims potentially bringing great social impact are fact-checked. When encountering seemingly suspicious claims without contextual information, regular users may struggle to seek related information by themselves: to look for any existing fact-checking results of a claim, they may not have a solid idea of which source has already reviewed the claim and end up randomly checking several websites (e.g., PolitiFact, Snopes, and FactCheck.org). In that regard, they may use Google Search, which could return excessive information, such as fact-checking reports and other unhelpful information, making it hard to resolve the concern. Another challenge is that if a claim has been fact-checked by multiple sources, their potentially different fact-checking verdict terminologies (e.g., “False,” “Mostly False,” “Misleading,” and “Mixed”) and rating criteria could result in confounding conclusions. For instance, regarding an identical false claim, a verdict could be rated “misleading” by one source while “mostly true” by another. Although some researchers compared the fact-checking results from a few popular sources and found them less likely to review the same claims \cite{ref29}, whether it is a general case for most sources is unclear. 

\begin{figure}[t]
  \centering
  \includegraphics[width=\textwidth]{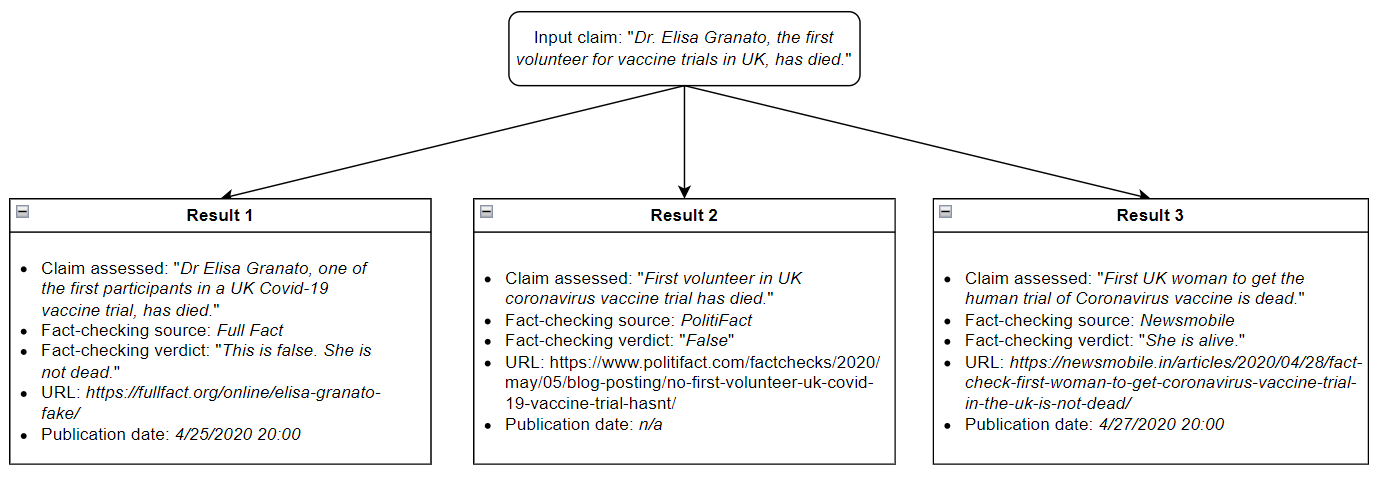}
  \caption{\centering An example introducing the structure of the results on Google Fact Check. Here, an input claim obtains three results, each containing a claim assessed by a source, a source name, a fact-checking verdict, a URL to the original webpage, and a publication date. Note that input claims may receive a different number of results.}
  \label{fig:result}
\end{figure}

A promising tool to address these two challenges (i.e., lack of useful tools to integrate fact-checking results and insufficient knowledge of the congruence of the fact-checking results rendered by various sources) is Google Fact Check Tools, a Google-based search engine for fact-checking results \cite{aboutGoogleFact}. Users can search for fact-checking results by complete claims or keywords. Each result on the results page contains a claim assessed by a source, a source name, a fact-checking verdict, a publication date, and a URL to the original report. Fig.~\ref{fig:result} shows the result structure in an example. An API is also available for software development, such as automated fact-checking. Although such a tool can significantly improve the efficiency of fact-checking, it is important to validate its ability to provide sufficient and reliable results before rolling it out. Besides, Google Fact Check is an ideal platform to compare the potentially varied fact-checking verdicts across sources, such as how many claims have been fact-checked by multiple sources and whether their fact-checking results are congruent. To the best of our knowledge, Google Fact Check is the only search engine specifically for fact-checking as of this study. Even though researchers contributed to addressing the aforementioned challenges by, for instance, indicating the limited ability of Google Fact Check to handle complex claims \cite{ref26} and different types of misinformation \cite{ref27} and analyzing the fact-checking results rendered by a few sources \cite{marietta2015fact,pereira2022characterizing}, no studies have thoroughly explored the performance of Google Fact Check or similar fact-checking-specific search engines, nor did they compare fact-checking results across sources on a general basis. Therefore, in this study, we aim to understand the usability of fact-checking-specific search tools by evaluating the performance of Google Fact Check regarding the quality of results in the dimensions that have not been fully explored, such as how relevant the retrieved fact-checking results are to the input claims and their potential correlations, so that we may shed light on the best practices for users to retrieve useful information. Specifically, we focused on the following questions:
\begin{enumerate}
    \item To what extent are the fact-checking results retrieved by Google Fact Check relevant to the input claims?
    \item Is there any correlation between the linguistic characteristics of input claims (e.g., length, emotional tone, analytical thinking level) and the retrieval of the best-matched fact-checking results?
    \item If multiple claims address the same issue but are described differently, to what extent are their fact-checking results congruent?
\end{enumerate}

In general, we made the following contributions:
\begin{enumerate}
    \item We evaluated the performance of Google Fact Check by analyzing the retrieved fact-checking results regarding 1,000 COVID-19-related false claims. In our experiments, 842 (84.2\%) of the claims did not retrieve any fact-checking results, and the remaining 158 (15.8\%) retrieved at least one result each. In total, 290 results were returned, among which 94.46\% were relevant to the input claims, i.e., the claims scrutinized by fact-checkers and the input claims addressed the same topic. Among these relevant results, 91.54\% were rated ``false'' or ``partly false'' by fact-checking sources with high reliability. 
    \item We explored the correlations between claims and fact-checking results via data analyses. The original claims receiving different verdicts (i.e., ``False,'' ``Partly False,'' ``True,'' and ``Unratable'') tend to reflect diverse emotional tones, and the claims scrutinized by different sources tend to be in various lengths and use dictionary words to various extents.
    \item We found that claim variations addressing the same issue are likely to obtain distinctive fact-checking results, shedding light on the best practices for users to retrieve the most useful information.
\end{enumerate}

The rest of this paper is organized as follows. In Section~\ref{sec:methodology}, we introduce our methodology, including the data collection process and data analysis methods. Then, we present our results in Section~\ref{sec:results}. Finally, we discuss our findings, limitations, and future work in Section~\ref{sec:conclusion}. 

\section{Methodology}
\label{sec:methodology}
\subsection{Data Collection}
\label{subsec:3_data_collection}
Misinformation can be found anywhere, and fact-checking can be applied to any topic. In this study, we decided to focus on COVID-19 as it was one of the representative topics arousing numerous rumors and conspiracies during the pandemic, even resulting in ``infodemic,'' \cite{diseases2020covid}, and there are sufficient datasets available for our experiments. We leveraged the FakeCovid \cite{shahi2020fakecovid}, a dataset consisting of over 5,000 COVID-19-related fact-checking results from Poynter and Snopes, two reputable fact-checking sources, where the former is also the leader of the International Fact-Checking Network (IFCN), the biggest fact-checking alliance joined by more than 100 fact-checking organizations globally \cite{ifcn_about_2024}. This dataset was compiled in the early stage of the pandemic when knowledge about COVID-19 was lacking, and related misinformation was rampant on social media. This multilingual dataset covers multiple domains of COVID-19, such as origin, spread, treatment, and conspiracy \cite{shahi2020fakecovid}. To study the performance of Google Fact Check, we excluded non-English results from the FakeCovid dataset and randomly selected 1,000 false claims for our experiments.


We programmed via Google Fact Check API to obtain the fact-checking results for these claims. The returned metadata were originally in JSON fields, including text, claimant, publisher, textual rating, and review date \cite{google_rest_claims_2024}. We renamed the fields to make them more readable: Input Claim, Claim Assessed by Fact-Checkers (i.e., the real claim fact-checked, which could be identical to or different from the input claim), Fact-Checking Source, Fact-Checking Verdict, Publication Date, and URL (see Fig.~\ref{fig:result} for an example). All the data we analyzed in this study were obtained in December 2022 via the API except the input claims from the FakeCovid dataset. The results are shown in Section~\ref{subsec:4_number}.

\subsection{Data Sanitization}
\label{subsec:3_data_sanitization}
\subsubsection{Relevance of Fact-Checking Results to Input Claims}
\label{subsec:3_relevance}
Since Google Fact Check is essentially a search engine, a claim assessed by fact-checkers could overlap with the input claim to different extents: (1) they are identical, i.e., the exact claim has been fact-checked; (2) the input claim has yet to be reviewed, but some other relatable claims are returned, so the relevance of the returned item is worth an examination. For instance, when the claim “Queen Elizabeth II is infected with the new coronavirus” has not been fact-checked yet, the fact-checking results for another claim “Queen Elizabeth died because of the COVID-19 vaccine” could be returned, even though the latter deviates from the point “infection.” To investigate to what extent the returned results are relevant to their original input claims, we recruited three coders from our research team to individually rate whether each claim assessed by fact-checkers is relevant to the corresponding input claim, i.e., whether they address the same issue. As a result, 262 (90.66\%) of the 289 fact-checking results were rated unanimously by three coders (i.e., ``relevant'' or ``irrelevant''), indicating a high agreement among coders, even though the Krippendorff’s alpha is \textit{K} = 0.476 \cite{krippendorff2018content}. Based on that, we rated a result ``relevant'' to the input claim if at least two of the three coders agreed. The rating results are shown in Section~\ref{subsec:4_number}.

\subsubsection{Mapping of Fact-Checking Verdict Terminology}
\label{subsec:3_mapping}
Each fact-checking result on Google Fact Check contains a verdict indicating the accuracy of a claim. Since Google Fact Check is a search engine collecting fact-checking results from a wide range of sources adopting different verdict terminologies (e.g., “Mostly False,” “Mixture,” and “Mostly Inaccurate”) and rating criteria, synonymous verdicts (e.g., “Mostly False” and “Partly True”) could bring a challenge to our analysis. We reviewed the verdict definitions adopted by all the sources involved in the returned results and summarized them as follows: if there is clear evidence, the verdict is essentially ``false,'' ``true,'' or a mixture of both; otherwise, the accuracy is inconclusive. Thus, we mapped the original verdicts into four categories: “False,” “Partly False,” “True,” and “Unratable.” Specifically, we directly mapped four verdicts (i.e., “False,” “Partly False,” “True,” and “Unratable”) to the categories under the same name and the other verdicts to these four categories based on the definitions adopted by their sources.\footnote{The complete datasets and documentation are available on OSF: \url{https://osf.io/zkbd4/?view_only=804bdd91c7f340eeacc3bf3494c06930}\label{fn}}

For the verdicts in a long sentence instead of simple words, the three coders in Section~\ref{subsec:3_relevance} manually reviewed and mapped them to the proper categories the same way they coded in Section~\ref{subsec:3_relevance}. The resulting Krippendorff’s alpha is \textit{K} = 0.817, indicating a high inter-coder agreement \cite{krippendorff2018content}. For any disagreements among the coders, we accepted the choice by at least two coders; if they all chose different categories, they had an open discussion until reaching an agreement. As a result, all the original verdicts were mapped to these four categories for better analysis. Although it is impossible to cover all the fact-checking sources in the world and the verdict terminologies they adopted, these four categories can theoretically cover any verdict from any source. The analysis results of the fact-checking verdicts are shown in Section~\ref{subsec:4_verdict}.

\subsection{Source Reliability}
\label{subsec:3_source}
The reliability of fact-checking sources is critical as biased sources are more likely to render unreliable fact-checking results, mislead the audience, and exacerbate the consequence of misinformation \cite{doi:10.1111/j.1468-2508.2005.00349.x}. We first investigated the frequency of each source being referenced by the fact-checking results we obtained, then we looked up the source reliability on two websites for media evaluation: Interactive Media Bias Chart by Ad Fontes Media \cite{adfontesmedia_interactive_2024} and Media Bias/Fact Check \cite{mediabiasfactcheck_2024}. They rated media sources from two dimensions: (1) reliability, i.e., how reliable the information from a source is, and (2) political leaning, i.e., whether the political leaning of a source is relatively neutral, left, or right. Since these two websites adopted different rating terminologies and criteria, similar to the approach introduced in Section~\ref{subsec:3_mapping}, we mapped the original reliability ratings to four categories: ``Trustworthy,'' ``Relatively Trustworthy,'' ``Relatively Untrustworthy,'' and ``Untrustworthy,'' and the original political leaning ratings to three categories: ``Left,'' ``Center,'' and ``Right.''\footref{fn} The analysis results of source reliability are presented in Section~\ref{4_source}.

\subsection{Correlation Between Claims and Results}
\label{subsec:3_relationship}
Understanding how the linguistic characteristics of input claims are likely to influence the quality of results may help us comprehend the best practices for getting the most useful results. We investigated the relationships between input claims and fact-checking results. Specifically, we leveraged LIWC, a software application for linguistic analysis, to quantify the linguistic characteristics of the 1,000 input claims, including word count, analytical thinking, clout, authentic, emotional tone, words per sentence, and dictionary words. As for the fact-checking results returned by Google Fact Check, we leveraged the dimensions analyzed in the previous sections, i.e., the number and relevance of fact-checking results, fact-checking verdicts, and source reliability. Then, we performed the following data analyses in Stata.

\subsubsection{Number of Results vs. Characteristics of Input Claims}
\label{subsec:3_no_input}
We tested whether the linguistic characteristics of input claims are likely to result in obtaining different numbers of fact-checking results. Since the variables do not follow normal distributions, we calculated Spearman’s rank correlation coefficients ranging from -1 to 1, where the positive and negative signs indicate the directions of correlation and a greater deviation from 0 indicates a stronger correlation \cite{spearman1961proof}. The results are shown in Section~\ref{subsec:4_input_number}.

\subsubsection{Relevance of Results vs. Characteristics of Input Claims}
\label{subsec:3_relevance_input}
In Section~\ref{subsec:3_relevance}, three coders reviewed the fact-checking results and evaluated whether each claim assessed by fact-checkers was relevant to the original input claim, i.e., whether they focused on the same topic. We wondered whether such relevance is related to the characteristics of input claims. To do so, we grouped the fact-checking results we obtained based on whether they were rated relevant or irrelevant in Section~\ref{subsec:3_relevance} and performed the Kruskal-Wallis H test for any significant difference in the characteristics of input claims. The results are shown in Section~\ref{subsec:4_input_relevance}. We performed the Kruskal-Wallis H test because the ``relevance'' is nominal data and should not be calculated as meaningful numbers. It is the same for the calculations in the following Sections~\ref{subsec:3_verdict_input} and \ref{subsec:3_source_input}.

\subsubsection{Fact-Checking Verdicts vs. Characteristics of Input Claims}
\label{subsec:3_verdict_input}
In Section~\ref{subsec:3_mapping}, the fact-checking verdicts from various sources were mapped to four categories (i.e., “False,” “Partly False,” “True,” and “Unratable”) for better analysis. We wondered whether claims with different characteristics are likely to obtain varied verdicts, e.g., claims showing a lower critical thinking level are more likely to be rated ``False.'' To do so, we first excluded the irrelevant and non-English fact-checking results coded in Section~\ref{subsec:3_relevance}. Then, we grouped the remaining results based on the four categories and performed the Kruskal-Wallis H test to detect any significant difference in the characteristics of input claims. The results are shown in Section~\ref{subsec:4_input_verdict}. 

\subsubsection{Fact-Checking Sources vs. Characteristics of Input Claims}
\label{subsec:3_source_input}
In Section~\ref{subsec:3_source}, we evaluated the reliability of the fact-checking sources. Some sources may have a taste for specific topics or claims in specific linguistic patterns, which could drive them to adopt various fact-checking methodologies and criteria. To better understand the potential relationship between sources and claim characteristics, we performed the Kruskal-Wallis H test to detect the significant difference in the characteristics of claims across the sources. The results are presented in Section~\ref{subsec:4_input_source}.

\subsection{Input Claim Variation}
\label{subsec:3_input_variation}
Even though we ensured the false claims leveraged in this study were not repeated, we noticed that some claims addressed the same issue in varied descriptions. For instance, the claims “Doctors in Japan advise people to drink water every 15 minutes to prevent an infection” and “Drinking water every 15 minutes will protect people from coronavirus” both addressed that drinking water is a promising treatment for COVID-19 but described differently. This observation is relatable to the real-life situation where a claim seems suspicious to the audience who may verify it by searching for any related information using their own words. Therefore, we also investigated how fact-checking results can be influenced by descriptive variations. 

To collect the varied claims addressing the same issue, we involved the coders who contributed in Sections~\ref{subsec:3_data_sanitization} and \ref{subsec:3_source}. Each coder was assigned a datasheet copy listing the fact-checking results rated ``relevant'' in Section~\ref{subsec:3_relevance}. The coders first individually reviewed and tagged each input claim with one or more keywords that can summarize the key topic of the claim. Then, the claims with identical or similar tags were grouped by manually dragging the corresponding rows in the datasheet. We considered ``similar'' tags as well because coders may forget what tags they previously used when coding for more than 200 claims. For instance, if there is a claim “There is a relationship between COVID-19 and the 5G network” at the beginning of the datasheet, it could be tagged “COVID-19 and 5G”; after a long coding process, the coder may see another similar claim “Coronavirus cases linked to 5G rollout” at the end of the datasheet, but they may forget the tag attached to the previous claim and thus use a new tag “Coronavirus and 5G,” although these two claims should share the same tag and grouped because they essentially described the same story. Therefore, after the initial tagging, each coder reviewed, grouped, and merged any tags that essentially meant the same. Then, we compared the coding results from the three coders and had an open discussion to resolve any disagreements. Note that the goal of tagging is to find out the claims addressing the same issue yet described differently, so it is acceptable if the coders tagged claims with different keywords as long as they grouped the same claims. For instance, the two claims mentioned above could be tagged as “COVID-19 and 5G” by one coder and “Coronavirus and 5G” by another, but it is acceptable if they both found and grouped these two claims addressing the same issue. 

We focused on the tags involving at least two input claims discussing the same issue yet in different descriptions and analyzed the congruence of their fact-checking results to explore how likely they could obtain the same fact-checking results. To quantify such a congruence, we calculated the Jaccard index for the fact-checking results of every two input claims with the same tag by counting their overlapped results (intersection set) and the unique results in total (union set), then dividing the intersection set by the union set \cite{karabiber_jaccard_2024}. The equation is:

\begin{equation}
    J(S_1, S_2) = \frac{|S_1 \cap S_2|}{|S_1 \cup S_2|} = \frac{\text{Number of Results in the Intersection}}{\text{Number of Results in the Union}}\label{eq}
\end{equation}


Since the Jaccard index is normally used to calculate the congruence between two groups, we calculated the average Jaccard index of all the permutation pairs for the tags involving more than two input claims. For instance, if there are three claims, A, B, and C, with the same tag, we first calculated the Jaccard indices of A and B, A and C, and B and C, respectively, then calculated the average. The results are shown in Section~\ref{subsec:4_input_variation}.

\section{Results}
\label{sec:results}
\subsection{Statistics of Fact-Checking Results}
\label{subsec:4_number}
Among the 1,000 false claims we leveraged, 842 (84.2\%) failed to get any results, 101 (10.1\%) received one result each, and 57 (5.7\%) received two or more results. The total number of the results is 290. Table~\ref{tab:number} shows more detail. We also found that all the reviewed claims are unique, i.e., no sources reviewed the identical claims. This corresponds to the findings in the previous studies \cite{marietta2015fact, pereira2022characterizing}. 

\begin{table}[tb]
    \centering
    \caption{The number of fact-checking results received by a claim and the corresponding number of input claims involved. 158 (15.8\%) input claims obtained 290 fact-checking results in total from Google Fact Check.}
    \begin{tabularx}{\linewidth}{| >{\centering}m{0.37\linewidth} | >{\centering}m{0.2\linewidth} | >{\centering\arraybackslash}m{0.4\linewidth} |}
        \hline
        \raggedright \textbf{Number of Fact-Checking Results Obtained by Each Claim} & \raggedright \textbf{Number of Input Claims Involved} & \raggedright\arraybackslash \textbf{Percentage (Total Number of Input Claims = 1000)}\\
        \hline
        0 & 842 & 84.20\%\\
        \hline
        1 & 101 & 10.10\%\\
        \hline
        2 & 31 & 3.10\%\\
        \hline
        3 & 8 & 0.80\%\\
        \hline
        4 & 9 & 0.90\%\\
        \hline
        5 & 1 & 0.10\%\\
        \hline
        6 & 4 & 0.40\%\\
        \hline
        8 & 1 & 0.10\%\\
        \hline
        10 & 3 & 0.30\%\\
        \hline
    \end{tabularx}
    \label{tab:number}
\end{table}

As for the relevance of results, one result was not in English and thus was ignored; 273 (94.46\%) of the remaining 289 results were rated relevant to the input claims by three coders because their claims assessed by fact-checkers were in English and focused on the same topic as the corresponding input claims; the remaining 16 (5.54\%) were rated irrelevant because they deviated from the main topic of the input claims. Table~\ref{tab:relevance} presents the results. 

\begin{table}[b]
    \centering
    \caption{The distribution of the relevant and irrelevant results rated by three coders. Note that we excluded one non-English result from the 290 results returned by Google Fact Check, so the total number of valid results was 289.}
    \begin{tabularx}{\linewidth}{| >{\centering}m{0.16\linewidth} | >{\centering}m{0.3\linewidth} | >{\centering\arraybackslash}m{0.51\linewidth} |}
        \hline
        \textbf{Relevance} & \textbf{Number of Results} & \raggedright\arraybackslash \textbf{Percentage (Total Number of Valid Results = 289)}\\
        \hline
        Relevant & 273 & 94.46\%\\
        \hline
        Irrelevant & 16 & 5.54\%\\
        \hline
    \end{tabularx}
    \label{tab:relevance}
\end{table}

Even though most (94.46\%) of the results were rated relevant to the input claims, the lack of results regarding 84.2\% of the input claims may not well help users fact-check the issues they have concerns about. According to the About page of Google Fact Check, the results were provided spontaneously by fact-checkers, i.e., when publishing a fact-checking report, publishers can opt to attach a ClaimReview markup to make it detectable by a search engine \cite{aboutGoogleFact}. Therefore, a possible explanation is that more fact-checking results did exist somewhere online but were not detectable by Google Fact Check, of which the usability was thus limited. 

\subsection{Statistics of Fact-Checking Verdicts}
\label{subsec:4_verdict}
To analyze the fact-checking verdicts, we only considered the 273 fact-checking results rated relevant to the original input claims in the section above. We further excluded a result in which the fact-checking verdict was not in English, so the total number of results to analyze was reduced to 272. Table~\ref{tab:verdict} shows the distribution of fact-checking verdicts. 217 (79.78\%) verdicts were “False” and 32 (11.76\%) were “Partly False.” These two categories debunking inaccuracies with clear evidence accounted for 91.54\% in total. As for the remaining, 22 (8.09\%) were ``Unratable'' without conclusive evidence to prove true or false, and only one (0.37\%) was ``True.'' Since the claims leveraged were from the FakeCovid dataset and were manually scrutinized by editors, they are more likely to be problematic, influential, and thus worth being fact-checked. This may explain why most of the verdicts we obtained were negative. Even so, it is reasonable to assume that professionals have a higher sensitivity to hot topics and misinformation so that these fact-checked claims are likely to reflect public focus in general. 

\begin{table}[tb]
    \centering
    \caption{The distribution of fact-checking verdicts. We excluded 16 irrelevant results and one result whose fact-checking verdict was not in English, so the total number of results involved was reduced to 272.}
    \begin{tabularx}{\textwidth}{| >{\raggedright}m{0.22\textwidth} | >{\centering}m{0.24\textwidth} | >{\centering\arraybackslash}m{0.51\textwidth} |}
        \hline
        \raggedright \textbf{Mapped Fact-Checking Verdict} & \raggedright \textbf{Number of the Original Verdicts Involved} & \raggedright\arraybackslash \textbf{Percentage (Total Number of Valid Fact-Checking Results = 272)}\\
        \hline
        False & 217 & 79.78\%\\
        \hline
        Partly False & 32 & 11.76\%\\
        \hline
        True & 1 & 0.37\%\\
        \hline
        Unratable & 22 & 8.09\%\\
        \hline
    \end{tabularx}
    \label{tab:verdict}
\end{table}

\subsection{Statistics of Source Reliability}
\label{4_source}
Table~\ref{tab:source} lists the distribution of the fact-checking sources referenced by the 272 valid fact-checking results in the section above and their reliabilities and political leanings. The top five sources referenced the most are PolitiFact (n = 53, 19.49\%), AFP Fact Check (n = 47, 17.28\%), Full Fact (n = 26, 9.56\%), Health Feedback (n = 20, 7.35\%), and FactCheck (n = 16, 5.88\%). In total, they were referenced by 162 (59.56\%) results. 11 (4.04\%) results were returned without source information.

We referred to two websites, Interactive Media Bias Chart and Media Bias/Fact Check, to evaluate each source's reliability and political leaning. Eight (32\%) of the 25 sources involved were rated by both websites, six (24\%) by one only, and 11 (44\%) by neither one. Among the sources rated by at least one website (n = 14, 56\%), all were rated ``Trustworthy'' or ``Relatively Trustworthy,'' indicating high reliability of their fact-checking verdicts; as for political leaning, 14 (63.64\%) of the 22 ratings in the table are ``Center,'' indicating a relatively unbiased stance in general. 

\begin{table*}[tb]
    \centering
    \caption{The distribution of the fact-checking sources referenced by 272 valid fact-checking results.}
    \scriptsize
    \begin{tabularx}{\textwidth}{| >{\raggedright}m{0.15\textwidth} | >{\centering}m{0.1343\textwidth} | >{\centering}m{0.13\textwidth} | >{\centering}m{0.13\textwidth} | >{\centering}m{0.13\textwidth} | >{\centering}m{0.13\textwidth} | >{\centering\arraybackslash}m{0.13\textwidth} |}
        \hline
        \centering \textbf{Source} & \textbf{Number of Reference} & \raggedright \textbf{Percentage (Total Number of Results = 272)} & \raggedright \textbf{Reliability Rated by Interactive Media Bias Chart} & \raggedright \textbf{Reliability Rated by Media Bias/Fact Check} & \raggedright \textbf{Political Leaning Rated by Interactive Media Bias Chart} & \raggedright\arraybackslash \textbf{Political Leaning Rated by Media Bias/Fact Check}\\
        \hline
        PolitiFact & 53 & 19.49\% & Trustworthy & Trustworthy & Center & Left\\
        \hline
        AFP Fact Check & 47 & 17.28\% & n/a & Trustworthy & n/a & Center\\
        \hline
        Full Fact & 26 & 9.56\% & n/a & Trustworthy & n/a & Center\\
        \hline
        Health Feedback & 20 & 7.35\% & Trustworthy & Trustworthy & Center & Center\\
        \hline
        FactCheck & 16 & 5.88\% & Trustworthy & Trustworthy & Center & Center\\
        \hline
        Snopes & 14 & 5.15\% & Trustworthy & Relatively Trustworthy & Center & Left\\
        \hline
        FACTLY & 14 & 5.15\% & -- & -- & -- & --\\
        \hline
        Boom & 13 & 4.78\% & Trustworthy & Trustworthy & Center & Right\\
        \hline
        Lead Stories & 12 & 4.41\% & Trustworthy & Trustworthy & Center & Center\\
        \hline
        Alt News & 7 & 2.57\% & -- & Trustworthy & -- & Left\\
        \hline
        USA Today & 7 & 2.57\% & Trustworthy & Relatively Trustworthy & Center & Left\\
        \hline
        Ghana Fact & 4 & 1.47\% & -- & -- & -- & --\\
        \hline
        Australian Associated Press & 4 & 1.47\% & -- & Trustworthy & -- & Center\\
        \hline
        The Washington Post & 4 & 1.47\% & Relatively Trustworthy & Relatively Trustworthy & Left & Left\\
        \hline
        Check4Spam & 3 & 1.10\% & -- & -- & -- & --\\
        \hline
        Newsmeter & 3 & 1.10\% & -- & -- & -- & --\\
        \hline
        Newsmobile & 2 & 0.74\% & -- & -- & -- & --\\
        \hline
        Newschecker & 2 & 0.74\% & -- & -- & -- & --\\
        \hline
        The Quint & 2 & 0.74\% & -- & Trustworthy & -- & Left\\
        \hline
        THIP Media & 2 & 0.74\% & -- & -- & -- & --\\
        \hline
        The Journal & 2 & 0.74\% & -- & -- & -- & --\\
        \hline
        FactCheckHub & 1 & 0.37\% & -- & -- & -- & --\\
        \hline
        Namibia Fact Check & 1 & 0.37\% & -- & -- & -- & --\\
        \hline
        Africa Check & 1 & 0.37\% & -- & Trustworthy & -- & Center\\
        \hline
        FactRakers & 1 & 0.37\% & -- & -- & -- & --\\
        \hline
        (null) & 11 & 4.04\% & -- & -- & -- & --\\
        \hline
    \end{tabularx}
    \label{tab:source}
\end{table*}

\subsection{Correlation Between Claims and Results}
\subsubsection{Number of Results vs. Characteristics of Input Claims}
\label{subsec:4_input_number}
Table~\ref{tab:correlation} lists Spearman’s rank correlation coefficients between the characteristics of input claims and their corresponding numbers of fact-checking results. No significant correlation was detected as no coefficient was greater than 0.4 or smaller than -0.4, indicating a weak correlation \cite{akoglu2018user}.

\begin{table}[tb]
    \centering
    \caption{The Spearman’s rank correlation coefficients between the characteristics of the input claims and the number of fact-checking results.}
    \begin{tabularx}{\linewidth}{| >{\raggedright}m{0.26\linewidth} | >{\centering}m{0.31\linewidth} | >{\centering\arraybackslash}m{0.4012\linewidth} |}
        \hline
        \textbf{Characteristics of Input Claims} & \raggedright \textbf{Number of Results (for all 1,000 Claims)} & \raggedright\arraybackslash \textbf{Number of Results (for 158 Claims Obtaining Results)}\\
        \hline
        Word Count & -0.172 & -0.396 \\
        \hline
        Analytical Thinking & -0.169 & -0.057 \\
        \hline
        Clout & -0.052 & -0.143\\
        \hline
        Authentic & -0.023 & -0.149 \\
        \hline
        Emotional Tone & 0.045 & 0.091 \\
        \hline
        Words per Sentence & -0.163 & -0.383 \\
        \hline
        Dictionary Words & -0.073 & -0.294 \\
        \hline
    \end{tabularx}
    \label{tab:correlation}
\end{table}

\subsubsection{Relevance of Results vs. Characteristics of Input Claims}
\label{subsec:4_input_relevance}
We performed the Kruskal-Wallis H test to detect the relationship between the relevance of results and the characteristics of input claims. Table~\ref{tab:testing_relevance} shows the results. There was no \textit{p}-value less than 0.05, although the \textit{p}-value for emotional tone (\textit{p} = 0.051) was close to the critical value when we accepted tie raking. Therefore, no significant difference in the characteristics of input claims was observed.

\begin{table}[tb]
    \centering
    \caption{Kruskal-Wallis H test results for the characteristics of input claims and result relevance.}
    \begin{tabularx}{\linewidth}{|>{\raggedright}m{0.45\linewidth} | >{\centering}m{0.1\linewidth} | >{\centering}m{0.1\linewidth} | >{\centering}m{0.2027\linewidth} | >{\centering\arraybackslash}m{0.1\linewidth} |}
        \hline
        \textbf{Characteristics of Input Claims} & \textbf{chi2} & \textit{\textbf{p}} & \textbf{chi2 with ties} & \textit{\textbf{p}}\\
        \hline
        Word Count & 4.426 & 0.109 & 4.454 & 0.108\\
        \hline
        Analytical Thinking & 1.502 & 0.472 & 1.565 & 0.457\\
        \hline
        Clout & 1.322 & 0.516 & 1.467 & 0.480\\
        \hline
        Authentic & 1.879 & 0.391 & 1.98 & 0.372\\
        \hline
        Emotional Tone & 3.755 & 0.153 & 5.954 & 0.051\\
        \hline
        Words per Sentence & 4.084 & 0.130 & 4.112 & 0.128\\
        \hline
        Dictionary Words & 0.404 & 0.817 & 0.405 & 0.817\\
        \hline
    \end{tabularx}
    \label{tab:testing_relevance}
\end{table}

\subsubsection{Fact-Checking Verdicts vs. Characteristics of Input Claims}
\label{subsec:4_input_verdict}
We performed the Kruskal-Wallis H test between the characteristics of input claims and the fact-checking verdicts. As Table~\ref{tab:testing_verdict} shows, most of the \textit{p}-values were greater than 0.05, indicating insignificant difference in the characteristics of input claims. The only significant result was observed for emotional tone (\textit{p} = 0.014) when tie ranking was adopted. 

\begin{table}[tb]
    \centering
    \caption{Kruskal-Wallis H test results for the characteristics of input claims and fact-checking verdicts.}
    \begin{tabularx}{\linewidth}{|>{\raggedright}m{0.45\linewidth} | >{\centering}m{0.1\linewidth} | >{\centering}m{0.1\linewidth} | >{\centering}m{0.2027\linewidth} | >{\centering\arraybackslash}m{0.1\linewidth} |}
        \hline
        \textbf{Characteristics of Input Claims} & \textbf{chi2} & \textit{\textbf{p}} & \textbf{chi2 with ties} & \textit{\textbf{p}}\\
        \hline
        Word Count & 3.326 & 0.344 & 3.347 & 0.341\\
        \hline
        Analytical Thinking & 3.145 & 0.370 & 3.266 & 0.352\\
        \hline
        Clout & 0.352 & 0.950 & 0.390 & 0.942\\
        \hline
        Authentic & 1.861 & 0.602 & 1.967 & 0.579\\
        \hline
        Emotional Tone & 6.627 & 0.085 & \textbf{10.574} & \textbf{0.014}\\
        \hline
        Words per Sentence & 4.301 & 0.231 & 4.329 & 0.228\\
        \hline
        Dictionary Words & 7.727 & 0.052 & 7.741 & 0.052\\
        \hline
    \end{tabularx}
    \label{tab:testing_verdict}
\end{table}

\subsubsection{Fact-Checking Sources vs. Characteristics of Input Claims}
\label{subsec:4_input_source}
We performed the Kruskal-Wallis H test for the characteristics of input claims based on fact-checking sources. The results are shown in Table~\ref{tab:testing_source}. We observed significant results for word count (\textit{p} = 0.035 without ties and \textit{p} = 0.034 with ties) and dictionary words (\textit{p} = 0.001 without ties and \textit{p} = 0 with ties), indicating that different sources tend to fact-check claims with varied lengths and using dictionary words at different extents. All the other \textit{p}-values were greater than 0.05 and thus insignificant. 

\begin{table}[tb]
    \centering
    \caption{Kruskal-Wallis H test results for the characteristics of input claims and fact-checking sources.}
    \begin{tabularx}{\linewidth}{|>{\raggedright}m{0.45\linewidth} | >{\centering}m{0.1\linewidth} | >{\centering}m{0.1\linewidth} | >{\centering}m{0.2027\linewidth} | >{\centering\arraybackslash}m{0.1\linewidth} |}
        \hline
        \textbf{Characteristics of Input Claims} & \textbf{chi2} & \textit{\textbf{p}} & \textbf{chi2 with ties} & \textit{\textbf{p}}\\
        \hline
        Word Count & \textbf{40.425} & \textbf{0.035} & \textbf{40.663} & \textbf{0.034}\\
        \hline
        Analytical Thinking & 27.519 & 0.383 & 28.576 & 0.331\\
        \hline
        Clout & 27.247 & 0.397 & 29.919 & 0.271\\
        \hline
        Authentic & 22.650 & 0.653 & 23.964 & 0.578\\
        \hline
        Emotional Tone & 13.917 & 0.974 & 21.851 & 0.697\\
        \hline
        Words per Sentence & 36.880 & 0.077 & 37.107 & 0.073\\
        \hline
        Dictionary Words & \textbf{56.741} & \textbf{0.001} & \textbf{56.841} & \textbf{0}\\
        \hline
    \end{tabularx}
    \label{tab:testing_source}
\end{table}

\subsection{Input Claim Variation}
\label{subsec:4_input_variation}
We calculated the Jaccard index to quantify the congruence of the fact-checking results regarding claim variations, i.e., claims addressing the same issue but in different descriptive ways. Table~\ref{tab:topics} shows the numbers of fact-checking results in union and intersection sets and the corresponding Jaccard indices regarding the 21 tags involving at least two claim variations. We calculated the regular Jaccard index for the 18 (85.71\%) tags involving two input claim variations and the averaged Jaccard index for the remaining three (14.29\%) tags involving three or more claim variations. 17 (80.95\%) tags achieved low similarities (not greater than 0.5). Even though three (14.29\%) tags achieved perfect similarity (100\%), they each only involved two claim variations that received one identical fact-checking result. Therefore, input claims in different descriptions are not likely to obtain identical fact-checking results. Even though descriptive variations tend to cover nuanced details that may produce different results, such a situation is relatable to real life: when people are concerned about the same issue, they may individually search for related information in their own words for verification. Therefore, our investigation regarding such an issue is insightful. 

\begin{table}[htbp]
    \centering  
    \caption{The Jaccard Index of each topic. }
    \scriptsize
    \begin{threeparttable}
        \begin{tabularx}{\textwidth}{| >{\raggedright}m{0.22\textwidth} | >{\centering}m{0.2\textwidth} | >{\centering}m{0.2\textwidth} | >{\centering}m{0.22\textwidth} | >{\centering\arraybackslash}m{0.1128\columnwidth} |}
            \hline
            \centering \textbf{Tag} & \raggedright \textbf{Number of Input Claims} & \raggedright \textbf{Number of Unique Results} & \raggedright \textbf{Number of Results in Common} & \raggedright\arraybackslash \textbf{Jaccard Index}\\
            \hline
            breath & 2 & 7 & 1 & 0.143\\\hline
            water & 2 & 2 & 1 & 0.5\\\hline
            garlic water & 2 & 5 & 3 & 0.6\\\hline
            water, vinegar & 5 & n/a & n/a & 0.233\tnote{a}\\\hline
            onion & 2 & 5 & 2 & 0.4\\\hline
            hydroxychloroquine & 2 & 2 & 0 & 0\\\hline
            smoking & 2 & 1 & 1 & 1\\\hline
            hand sanitizer & 2 & 2 & 2 & 1\\\hline
            5G & 2 & 4 & 0 & 0\\\hline
            5G equipment label & 2 & 2 & 0 & 0\\\hline
            lab product & 3 & n/a & n/a & 0.111\tnote{a}\\\hline
            airborne & 2 & 2 & 0 & 0\\\hline
            asymptomatic & 2 & 3 & 0 & 0\\\hline
            surgical mask color & 2 & 3 & 0 & 0\\\hline
            UK volunteer & 4 & n/a & n/a & 0.5\tnote{a}\\\hline
            Ronaldo & 2 & 4 & 1 & 0.25\\\hline
            Chales Lieber & 2 & 2 & 0 & 0\\\hline
            Trump positive & 2 & 2 & 2 & 1\\\hline
            China 20,000 patients & 2 & 4 & 1 & 0.25\\\hline
            Italy cure & 2 & 4 & 0 & 0\\\hline
            Tasuku Honjo & 2 & 8 & 2 & 0.25\\
            \hline
        \end{tabularx}
        \begin{tablenotes}
            \item[a] The Jaccard Index is the average of all comparison pairs due to the existence of more than two pairs to compare. 
        \end{tablenotes}
    \end{threeparttable}
    \label{tab:topics}
\end{table}

\section{Discussion and Conclusion}
\label{sec:conclusion}
In this study, we explored the promise of fact-checking-specific search engines by evaluating the performance of Google Fact Check, a Google-based search engine for fact-checking results and the only fact-checking-specific search engine as of this study. Our study was motivated by two practical issues. First, even though Google Fact Check and other similar tools bring great convenience to fact-checking tasks and have been leveraged as a core component of some automated fact-checking frameworks, it is necessary to validate its usability, such as whether the fact-checking results it renders are helpful and reliable. Second, it was unclear whether different fact-checking sources tend to fact-check the same claims and whether their verdicts tend to be congruent, and a search engine collecting fact-checking results from various sources could pave the way for an investigation. We retrieved the fact-checking results for 1,000 COVID-19-related false claims via Google Fact Check API and analyzed them from the perspectives of result quality and the correlations between input claims and fact-checking results. We found that most claims did not retrieve fact-checking results, even though the returned results were relatively relevant to the corresponding input claims and tended to be reliable. Furthermore, we did not detect significant correlations between the linguistic characteristics of input claims and the corresponding fact-checking results, except that different fact-checking sources are less likely to repeatedly fact-check identical content and tend to check the claims with various lengths and the usage of dictionary words. We also found that the variations of input claim wording are likely to result in different fact-checking information. Based on these findings, we suggest that the quantities of the fact-checking results rendered by Google Fact Check and similar tools can be optimized by, for instance, enhancing collaborations with fact-checking sources to broaden the result scope. Users may not necessarily worry about the discrepancies among sources regarding the same content, which is not likely to be repeatedly checked across sources; however, they could slightly adjust the input wording for more potential fact-checking results, although the linguistic characteristics of input claims are not likely to significantly influence the quality of results. 

There are some limitations that we could not bypass in this study. First, due to the large amount of input claims we leveraged (n = 1,000), we retrieved the fact-checking results via API programming. However, we noticed in our preliminary experiment that the results from the API may not be exactly the same as those on the results page. We did not delve into this observation because API programming is prevalent in software development, such as automated fact-checking frameworks. Considering that our results were compiled in December 2022, future work may continue to investigate whether such a difference should be a big concern and whether the performance of Google Fact Check has been significantly optimized since then. Second, we noticed that a small number of claims seem irrelevant to COVID-19 (e.g., ``Trump and McConnell are blocking stimulus checks for Americans married to immigrants''). However, we then found them still in the context of COVID-19. With the large amount of claims we leveraged (n = 1,000), they are less likely to compromise the validity of the study. That said, future work may leverage the claims more directly related to a topic. Third, the subjectivity in this study was unavoidable. Even though we recruited three coders who manually reviewed the retrieved results by following reasonable criteria, the accuracy of the coding results was not fully under control. For instance, in Section~\ref{subsec:4_verdict}, there was a ``true'' verdict, which should have been rated ``irrelevant'' in Section~\ref{subsec:3_relevance} and ignored because its claim assessed by fact-checkers (i.e., ``Disposable masks should always be worn colored-side-out'') was not addressing the same issue as the original input claim (i.e., ``Two ways of wearing a surgical mask: Colored side out if you are sick and white side out if you do not want to become sick''), that is, the verdict ``true'' did not indicate that the original claim was true but rather another irrelevant claim, but it was still rated ``relevant'' by two of the three coders and kept for data analysis thereafter. As for the evaluation of source reliability, even though we referenced two professional websites for media evaluation, as emphasized on their About pages, their ratings could also be more or less biased. Although subjectivity is unavoidable by nature in any qualitative research, future work may recruit more coders and leverage more rating sources and criteria to further reduce the potential influence of subjectivity. Finally, as we introduced in \ref{subsec:3_data_collection}, even though fact-checking is applicable to any topic, this study centered on COVID-19-related misinformation due to its extensive rampancy, the severe consequence, and abundant datasets. Future studies may extend this study to other topics, such as politics, climate change, and influential social problems.




\bibliographystyle{splncs04}
\bibliography{ref}

\end{document}